\documentclass[sigplan,nonacm,screen]{acmart}

\usepackage{float}
\usepackage{algorithm}
\usepackage{algpseudocode}
\usepackage{amsmath}
\usepackage{todonotes}
\usepackage{csquotes}
\usepackage{tikz}
\usepackage{pdfpages}
\usepackage{subcaption}
\usepackage{graphicx}
\usepackage{placeins}

\AtBeginDocument{%
  \providecommand\BibTeX{{%
    \normalfont B\kern-0.5em{\scshape i\kern-0.25em b}\kern-0.8em\TeX}}}

\newcommand*\circled[1]{\tikz[baseline=(char.base)]{\node[shape=circle,draw,inner sep=1pt] (char) {#1};}}

\setcopyright{none}
\copyrightyear{2021}
\acmYear{2021}

\begin{document}

\title{\textsc{Bristle}: Decentralized Federated Learning in Byzantine, Non-i.i.d. Environments}

\author{Joost Verbraeken}
\email{j.verbraeken@student.tudelft.nl}
\affiliation{%
  \institution{Delft University of Technology}
  \city{Delft}
  \country{The Netherlands}
}

\author{Martijn de Vos}
\email{M.A.deVos-1@tudelft.nl}
\affiliation{%
  \institution{Delft University of Technology}
  \city{Delft}
  \country{The Netherlands}
}

\author{Johan Pouwelse}
\email{J.A.Pouwelse@tudelft.nl}
\affiliation{%
  \institution{Delft University of Technology}
  \city{Delft}
  \country{The Netherlands}
}

\begin{abstract}
Federated learning (FL) is a privacy-friendly type of machine learning where devices locally train a model on their private data and typically communicate model updates with a server.
In decentralized FL (DFL), peers communicate model updates with each other instead.
However, DFL is challenging since (1) the training data possessed by different peers is often non-i.i.d. (i.e., distributed differently between the peers) and (2) malicious, or Byzantine, attackers can share arbitrary model updates with other peers to subvert the training process.

We address these two challenges and present \emph{Bristle}, middleware between the learning application and the decentralized network layer.
Bristle leverages transfer learning to predetermine and freeze the non-output layers of a neural network, significantly speeding up model training and lowering communication costs.
To securely update the output layer with model updates from other peers, we design a fast distance-based prioritizer and a novel performance-based integrator.
Their combined effect results in high resilience to Byzantine attackers and the ability to handle non-i.i.d. classes.

We empirically show that Bristle converges to a consistent 95\% accuracy in Byzantine environments, outperforming all evaluated baselines. In non-Byzantine environments, Bristle requires 83\% fewer iterations to achieve 90\% accuracy compared to state-of-the-art methods. We show that when the training classes are non-i.i.d., Bristle significantly outperforms the accuracy of the most Byzantine-resilient baselines by 2.3x while reducing communication costs by 90\%.
\end{abstract}



\maketitle

\section{Introduction}

Machine learning applications have gained significant traction and are widely used for various purposes, such as understanding consumer preferences \cite{chapelle2005machine}, recognizing pictures \cite{krizhevsky2012imagenet}, predicting keystrokes \cite{RN8}, or translating texts~\cite{chui2018notes}.
Traditionally, these applications use neural networks trained on a single server using a tremendous amount of data generated by a large number of geo-distributed, heterogeneous edge devices such as smartphones, IoT devices, or autonomous vehicles~\cite{hu2020distributed}.
However, centralized training on data streams generated by such devices is limited by the following three factors.
First, transmitting training data over the Internet can pose a significant burden on backbone networks.
This burden is particularly problematic when media such as photos or videos are used as training data and is worsened by the fact that most learning applications communicate with cloud providers over wireless links~\cite{RN1007,RN1008}.
Second, maintaining a central server architecture can quickly become expensive and time-consuming as the number of devices increases~\cite{durkee2010cloud}.
Third, transmitting personal and sensitive information over the Internet, such as text conversations or photos, to cloud providers raises privacy concerns and is in certain jurisdictions not even allowed by regulations such as the US HIPAA~\cite{RN261} and the European GDPR law~\cite{RN262}.

\begin{figure}
    \includegraphics[width=\linewidth]{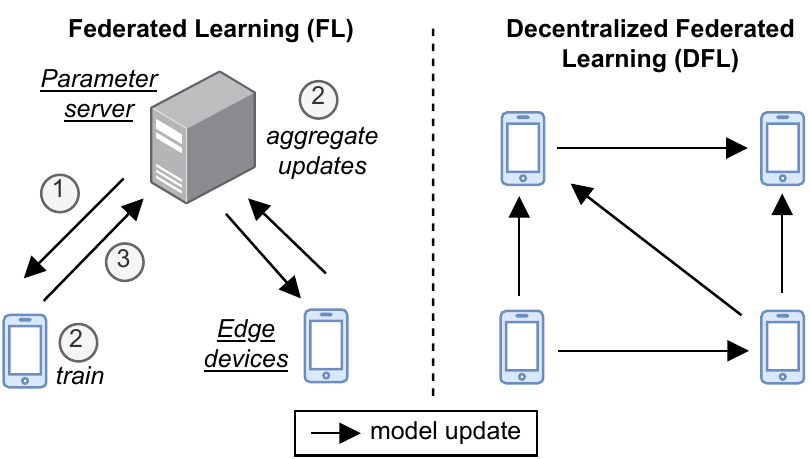}
    \caption{Federated learning using a parameter server (left) and decentralized federated learning (right).}
    \label{fig:architectures}
\end{figure}

\emph{Federated Learning (FL)} overcomes these three limitations.
With FL, edge devices do not send their data to the server training the model but instead communicate model updates to a so-called \emph{parameter server}, also see the left side of Figure~\ref{fig:architectures}~\cite{RN116}.
The parameter server coordinates the learning process by pushing model updates to devices (step~\circled{1}), and peers train the model with their private data on-device (step~\circled{2}).
Then they send the updated model back to the parameter server (step~\circled{3}), after which the server \emph{aggregates} the incoming model updates into a global model (step~\circled{4}).
FL sidesteps the need for the data to leave the device, improving privacy, lowering the computational burden for the server, and decreasing the communication overhead when the model update is smaller than the data to be transmitted per iteration.
FL is nowadays used for various applications, including next-word prediction on keyboards~\cite{RN154,RN207,RN152,RN9}, speech recognition~\cite{RN209}, wireless communications~\cite{RN158, RN157}, human activity recognition~\cite{RN160}, and health applications~\cite{RN210, RN170, RN212, RN169}.

The centralized architecture shown in Figure~\ref{fig:architectures} is typical for FL systems.
However, even though the parameter server synchronizes the learning process between remote edge devices, this approach comes with significant drawbacks~\cite{RN114, RN123, RN124}.
Notably, the parameter server not only poses a single point of failure susceptible to crashes or hacks, but it may also become a performance bottleneck as the number of devices pushing model updates increases~\cite{RN123}.
These issues motivate further research to remove dependency on the parameter server and train models in a decentralized manner~\cite{RN123, RN125, RN126, RN127}.
With \emph{decentralized federated learning} (DFL), each peer is connected to a subset of the other peers in the network from which it receives incoming models and to which it pushes its updated models~\cite{RN128}.
We visualize this approach on the right side of Figure~\ref{fig:architectures}.

DFL has resulted in a new wave of learning methods that achieve comparable model accuracy as state-of-the-art FL approaches~\cite{RN377} while boasting several significant advantages.
First, decentralized networks are fault-tolerant by design since peers continuously update their knowledge about which other peers are online or stopped communicating\cite{wallach2002survey}.
Second, decentralized networks are self-scaling. When a new peer joins the network, other peers can start communicating with the newly-joined peer seamlessly.
Because there is usually a maximum number of other peers with which each peer communicates, the connection ratio will drop, but the network will typically still be strongly connected \cite{felber2004self}.
Third, DFL unlocks AI with zero-cost infrastructure (from the developer's perspective) since the computational power needed to train the network is delivered by all peers in the network working together instead of a parameter server under centralized control \cite{endler2011defining}.

\textbf{Limitations of DFL.}
Despite the benefits of DFL, we identify three challenges that reduce its potential for practical applications.
The first barrier is that both existing FL and DFL architectures have to consider \emph{Byzantine attacks}, the situation where malicious peers aim to undermine the model's training by purposefully sending specific model updates to the parameter server or other peers~\cite{RN11}.
Since malicious peers keep their data private, they can easily \enquote{poison} their model and disseminate this malicious model without repercussions.
Byzantine attacks are often addressed by the Gradient Aggregation Rule (GAR), which aggregates and combines incoming model updates~\cite{RN79, RN11, RN12}.
At the same time, state-of-the-art GARs typically assume that the majority of peers act honestly, which is hard to guarantee in networks with open participation~\cite{RN1000}.

The second challenge is that the performance of conventional FL systems degrades significantly when being deployed in an environment where the distribution of the training data differs between peers.
This is also known as \emph{non-i.i.d.} (not independent and identically distributed) data.
Given that data is typically non-i.i.d. in an FL environment~\cite{RN12, RN8, RN11}, dealing with non-i.i.d. data is considered a \textit{key challenge} in FL~\cite{RN7}.
The inability of most GARs to handle non-i.i.d. data results in Catastrophic Forgetting, a phenomenon where the peers overwrite model parameters that were important to predict a particular class distribution with parameters essential to predict another class distribution~\cite{RN283}.

The third challenge relates to the communication requirements to disseminate model updates among peers in the network.
In DFL, each peer disseminates a copy of the current model at every training iteration to several other peers.
However, modern neural networks may consist of millions of parameters~\cite{RN1009, rn1010}, sometimes requiring gigabytes of data to be transferred for each model exchange.
Facilitating this amount of data quickly becomes infeasible when the number of devices and the frequency of model updates increase~\cite{sattler2020trends}.

Although some DFL systems are to a certain extent Byzantine-resilient~\cite{RN12, RN79, RN131, RN132}, able to deal with non-i.i.d. data~\cite{RN253, RN116, RN254}, or focus on reducing the communication costs~\cite{RN351, RN331, RN332}, there exists - to the best of the author's knowledge - no DFL solution that addresses all three challenges simultaneously.
We argue that such a solution is a key requirement for a wide deployment of DFL systems.

\textbf{Our Contributions. }
To the best of the author's knowledge \cite{RN149,RN130,RN131,RN132}, only four papers have ever been published about DFL that are Byzantine-resilient (first challenge), none of which are suitable for situations where the data is non-i.i.d. (second challenge) or decrease the communication requirements (third challenge).
Therefore, we present \emph{Bristle}: the first Byzantine-resilient and communication-efficient approach for DFL in environments with non-i.i.d. classes.\footnote{\texttt{Bristle} is an acronym for "Byzantine-Resilient mIddleware for StochasTic federated LEarning".}
Bristle is a pure edge solution and acts as a middleware between the machine learning application and the decentralized network layer by combining three techniques:
\begin{enumerate}
    \item Using \emph{deep transfer learning}, model training with Bristle converges quickly and achieves high accuracy with low communication overhead and while requiring low amounts of on-device data (Section~\ref{sec:transferlearning}).
    \item With a fast \emph{distance-based prioritizing step} during model aggregation based on an explore-exploit trade-off, Bristle pre-filters incoming model updates that are estimated to improve the current model (Section~\ref{sec:distancebasedprioritizer}).
    \item With a novel \emph{performance-based integrator}, Bristle provides state-of-the-art Byzantine-resilience, even when the classes are unevenly spread across peers (non-i.i.d.). Unlike related solutions, this component performs per-class performance measurements instead of per-model measurements (Section~\ref{sec:performancebasedintegrator}).
\end{enumerate}
Similar to related work in the FL domain, Bristle focuses on supervised learning problems that train a neural network for classification~\cite{RN79, RN6, RN91}.
We implement all elements of Bristle as an Android library and open-source our implementation (see Section~\ref{sec:implementation}).

\begin{figure}[t]
    \includegraphics[width=0.7\linewidth]{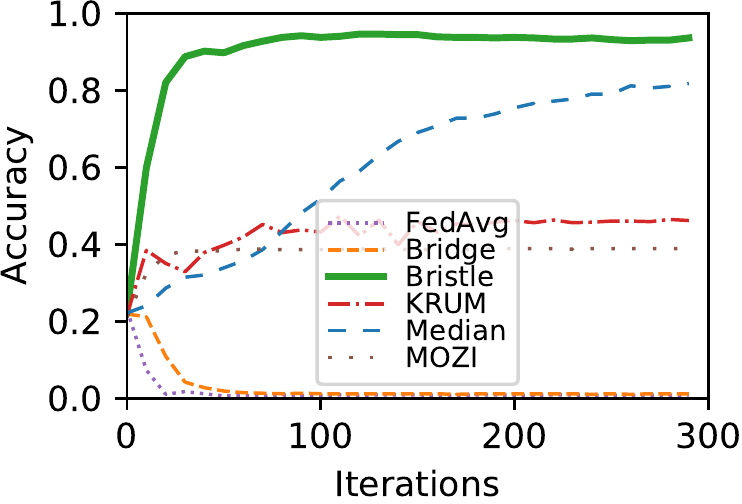}
    \caption{Our Bristle middleware for decentralized federated learning outperforms state-of-the-art solutions in a Byzantine and non-i.d.d. environment.}
    \label{fig:intro}
\end{figure}

With an extensive set of experiments with the popular MNIST dataset, we evaluate the performance and resilience of Bristle.
We compare Bristle with five related approaches for FL and quantify the Byzantine-resilience of these approaches under four attacks in the domain of FL, aimed at reducing the model's accuracy.
The experiments show that even in highly Byzantine environments where the classes are non-i.i.d., Bristle not only withstands all evaluated attacks but also outperforms all related approaches in terms of convergence speed, accuracy, consistency, and communication efficiency.
We highlight one of the key results from our experiments in Figure~\ref{fig:intro}, showing model accuracy as peers train the model.
In this figure, the performance of Bristle and evaluated DFL approaches is illustrated in a Byzantine (40\% of the peers perform a label-flip attack) and non-i.i.d. environment (the class overlap between the peers is just 40\%).
Despite this challenging environment, Bristle quickly converges and outperforms all other approaches in terms of accuracy while reducing communication costs by over 90\%.

\begin{figure*}[t]
    \includegraphics[width=\textwidth]{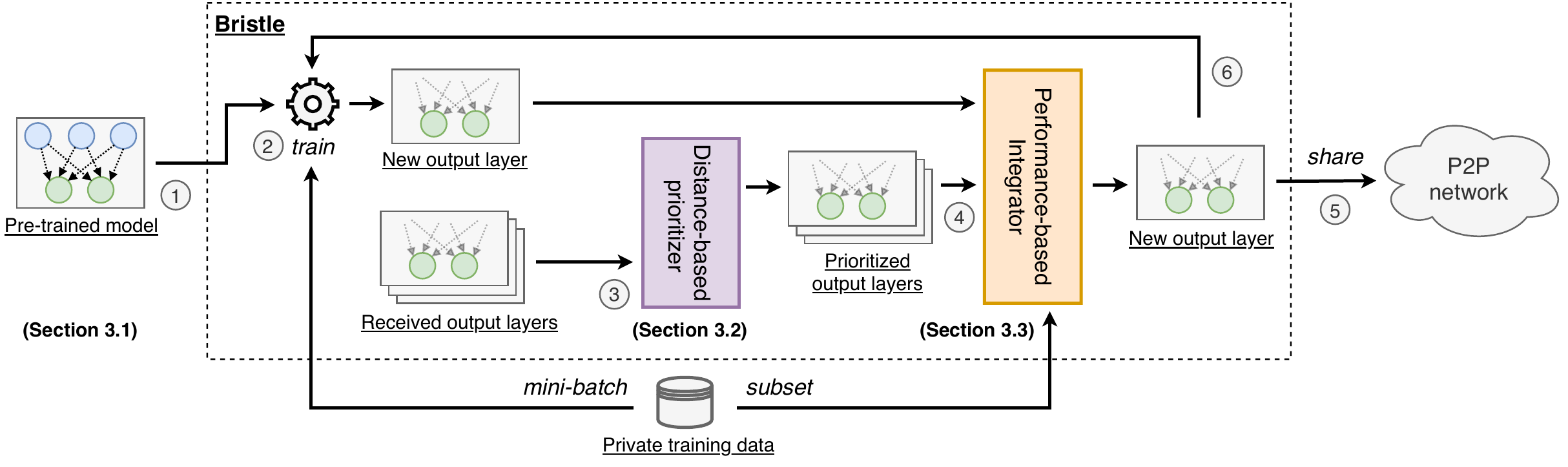}
    \caption{High-level overview of Bristle, our middleware for decentralized federated learning.}
    \label{fig:bristle}
\end{figure*}

\section{System and Threat model}
We now describe our system and threat model, and state the assumptions underlying our work.

\textbf{Network Model.}
We consider an unstructured, strongly connected peer-to-peer network with $n$ peers.
Each peer knows the network addresses of other peers in the network.
We assume unreliable and unordered network channels between peers.
In addition, we do not place any restriction on peers joining the system, and peers can join or leave the system at any time.
We consider the mitigation of attacks targeted at the network layer, e.g., the Eclipse Attack, beyond the scope of this work.

\textbf{Training Data.}
Each participating peer $ i $ acts on local training dataset $D_i$ which size is denoted by $|D_i|$.
Training data never leaves the device, and $ D_i $ is only known to peer $ i $.
In contrast to most related work, we assume that the number of samples per class is not distributed uniformly among peers (non-i.i.d), which is a key characteristic of FL environments \cite{RN11, RN12, RN8}.
However, we also assume that for each class the samples are distributed uniformly (i.i.d.) between peers as a prerequisite for the performance-based integrator (see Section~\ref{sec:performancebasedintegrator}).

\textbf{Model Training.}
Our approach focuses on supervised classification problems where the dataset consists of a collection of input-output pairs.
The input is a chunk of data with a fixed size, and the output is a qualitative label associated with a class.
The devices aim to minimize the loss function by tuning the parameters in the neural network such that its predictive capabilities are maximized.
We use the negative log-likelihood of the ground truth class as the loss function.
Since this problem is intractable for complex models, we use, in line with most literature \cite{RN190, RN11, RN194, RN195}, a technique called Gradient Descent (GD) which iteratively takes the derivative of the loss function with respect to the training data and then moves the hyperparameters in the direction of the gradient.
However, because the local dataset can be large, it can take a long time for GD to converge~\cite{RN1011}.
Therefore, we use the faster Stochastic Gradient Descent (SGD), where a subset (mini-batch) of five samples is stochastically sampled from the dataset to update the parameters in a particular iteration.

\textbf{Threat Model.}
Our work assumes an environment with \emph{an unconstrained number of Byzantine attackers} aiming to subvert the model's performance.
This includes the Sybil Attack, an attack where a malicious peer joins the network under many different identities to prevent model convergence~\cite{RN39}.
It also includes collusion, the situation where malicious peers in concert attempt to undermine model convergence.
Byzantine attackers can send arbitrary model updates to any peer in the network, but have no access to the private samples of benign peers.

\section{Bristle: Middleware for Decentralized Federated Learning}
Bristle enables DFL that can handle non-i.i.d. classes and thwart Byzantine attacks while improving communication costs compared to existing approaches.
We provide a high-level overview of Bristle and the actions performed during a training iteration in Figure~\ref{fig:bristle}.
First, Bristle takes advantage of deep transfer learning to determine the non-output layers of each node before the first training iteration (step~\circled{1}).
As we will experimentally demonstrate in Section~\ref{sec:results}, this significantly reduces communication costs compared to communicating the entire model, speeds up the model convergence rate, and acts as a prerequisite to learning on non-i.i.d. classes.

The main innovation of Bristle is its two-phased GAR that integrates received models.
During each training iteration, a peer $ i $ trains the local model on a mini-batch of their private dataset $ D_i $ and only updates the output layers (step~\circled{2}).
It then forwards the updated output layer\footnote{We use the terms output layer and model interchangeably in this work.} and the output layers it has received from other peers to a distance-based prioritizer (step~\circled{3}).
This fast distance-based prioritizer estimates the best candidates for integration based on an explore-exploit ratio and forwards them to the performance-based integrator (step~\circled{4}).
The performance-based integration is more computationally demanding and integrates the prioritized output layers into the peer's current output layer.
This integration process is both Byzantine-resilient and capable of continual learning, facilitated by selective updating of the output layer, per-class performance evaluations, and a carefully crafted weighted averaging function.
Our main motivation to use a two-staged approach is that a distance-based prioritizer on itself is ineffective and performance-based integration on all received models is too computationally expensive since it requires the evaluation of incoming models on private data.
Finally, Bristle transmits the output layer to a few connected peers (step~\circled{5}) and uses the new output layer as input for the next training iteration (step~\circled{6}).
In the remainder of this section, we elaborate on the model pre-training, the distance-based prioritizer, and the performance-based integrator.

\subsection{Bootstrapping Bristle with Transfer Learning}\label{sec:transferlearning}
In conventional FL settings, the parameters of the entire neural network are shared with the parameter server or, when using DFL, with other peers.
These parameters include all weights and biases of the input layer, hidden layers, and the output layer.
In Bristle, we avoid exchanging the full neural network between peers.
Instead, we leverage a popular ML technique called \emph{deep transfer learning} which re-uses a neural network that has been trained on another dataset with comparable low-level features as the training data (step \circled{1} in Figure~\ref{fig:bristle})~\cite{tan2018survey}.
More specifically, we copy and subsequently freeze the non-output layers from another model.
Bristle then only trains and exchanges the output layer, which has three major advantages:
\begin{enumerate}
    \item Copying the non-output layers from a well-trained model significantly \emph{improves the convergence rate} when training the model in a decentralized fashion on devices with lower hardware capabilities.
    \item Freezing non-output layers \emph{reduces communication overhead} since only the output layers have to be shared among peers.
    \item Freezing the non-output layers is a key step towards \emph{continual learning}, which further increases the performance and robustness of Bristle (see Section~\ref{sec:performancebasedintegrator}).
\end{enumerate}

To bootstrap Bristle with a pre-trained model, we must assume that for the dataset we want to train on, there exists a vastly bigger dataset with roughly the same low-level features that we can use to pre-train a similar model.
Considering the extent to which transfer learning is nowadays used in practical learning problems, we argue that this assumption is realistic and does not prohibitively degrade the applicability of Bristle~\cite{RN324,pan2010feature}.
The training of the initial neural network can be performed offline by system designers or volunteers.
The pre-trained model can then be shared with peers by bundling them in the (mobile) application or be served by a trusted server that sends interested peers the pre-trained model upon request.
We assume that the developer knows the number of classes beforehand to determine the size of the output layer.

\subsection{Distance-based prioritizer (DBP)}\label{sec:distancebasedprioritizer}
\begin{figure}[t]
    \includegraphics[width=0.7\linewidth]{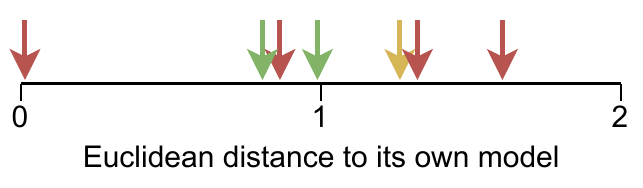}
    \caption{Average Euclidean distance from a given model to benign (green) and malicious (red) models, and models trained on an entirely different data distribution (orange).}
    \label{fig:ratios}
\end{figure}

We now elaborate on the distance-based prioritizer (DBP) in Bristle (step \circled{2} in Figure~\ref{fig:bristle}).
The inputs to the DBP are the output layers received from other peers during the last training iteration.
To explain the motivation behind the DBP, we first examine Figure~\ref{fig:ratios} where the average Euclidean distance is shown from a non-i.i.d. model (configured as explained in Section~\ref{sec:noniiddata}) to two benign models (shown in green), a benign model that is trained on an entirely different data distribution (shown in orange), and several Byzantine models (shown in red).
The distances in this figure are based on the models exchanged during our experiments (see Section~\ref{sec:results}).
For each distance shown in Figure~\ref{fig:ratios}, we took the average distance over 100 runs (with the current and the other model both trained ten times on different training data) and trained all models to convergence before comparing the distances.
The figure shows that based on the Euclidean distance alone, we cannot reliably determine if a certain model is benign because, in this example, the distance to a Trimmed Mean attack model is between the distances to two benign models.
However, the distance may clearly help prioritizing certain models since models with rather low or high distances appear to be more likely to be malicious.

\begin{figure}[b]
    \includegraphics[width=0.7\linewidth]{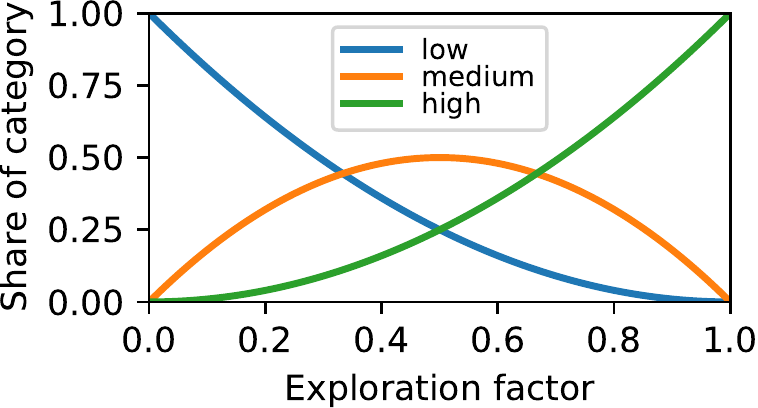}
    \caption{Proportion of models sampled by the distance-based prioritizer from each of the three categories, based on the exploration-exploitation ratio $\alpha$.}
    \label{fig:distancebasedweights}
\end{figure}

Based on this insight, we segment the received models into three equally-sized categories with a low, medium, and high distance from our current model.
Then, we uniformly sample models from each of these categories.
The number of models sampled from each category depends on the exploration-exploitation ratio $\alpha$ where $\alpha = 0$ exclusively samples low-distance models (exploitation-dominant) and $\alpha = 1$ exclusively samples high-distance models (exploration-dominant).
The optimal value of $\alpha$ depends on the maximum distance that benign models can reasonably have to each other, which is highly dependent on the degree of asynchrony (higher asynchrony $\rightarrow$ higher $\alpha$), the expected ratio of benign to Byzantine peers (more Byzantine peers $\rightarrow$ lower $\alpha$), and the degree of non-i.i.d.-ness (higher non-i.i.d.-ness $\rightarrow$ higher $\alpha$).
The advantage of using three categories is that in situations where the accuracy is clearly sub-optimal and a significant number of Byzantine attackers are present, the developer may want to emphasize integrating models with a medium distance since these will often perform best.
$ f_{l} $, $ f_{m} $ and $ f_{h} $ denote the fraction of samples from the low, medium and high distance categories, and are determined as follows:

\begin{equation}
    \begin{gathered}
    f_{l} = (1 - \alpha)^2, f_{m} = -2\alpha^2 + 2\alpha, f_{h} = \alpha^2\\
    \end{gathered}
\end{equation}

Note that $ f_{l} + f_{m} + f_{h} = 1 $.
Figure~\ref{fig:distancebasedweights} visualizes the relation between the above values and $ \alpha $.
We also parameterize the maximum number of models that pass our DBP, denoted by $ \beta $.
Since the runtime of our performance-based integrator scales with the input size, devices with lower performance can opt for a lower value for $ \beta $.

\subsection{Performance-based Integrator (PBI)}\label{sec:performancebasedintegrator}
The ability of our distance-based prioritizer to recognize Byzantine or \emph{stale} models that are trained on a lower number of iterations is rather limited.
Byzantine attacks are easy to launch in open networks, and stale models are a regular phenomenon in FL systems~\cite{li2020federated}.
To address these concerns, we assess the performance of each model that passes the distance-based prioritizer on a small \emph{test dataset} (step \circled{4} in Figure~\ref{fig:bristle}) using a performance-based integrator (PBI).
This test dataset is a random subset of the full private dataset owned by the peer.

\begin{figure}[b]
    \includegraphics[width=\linewidth]{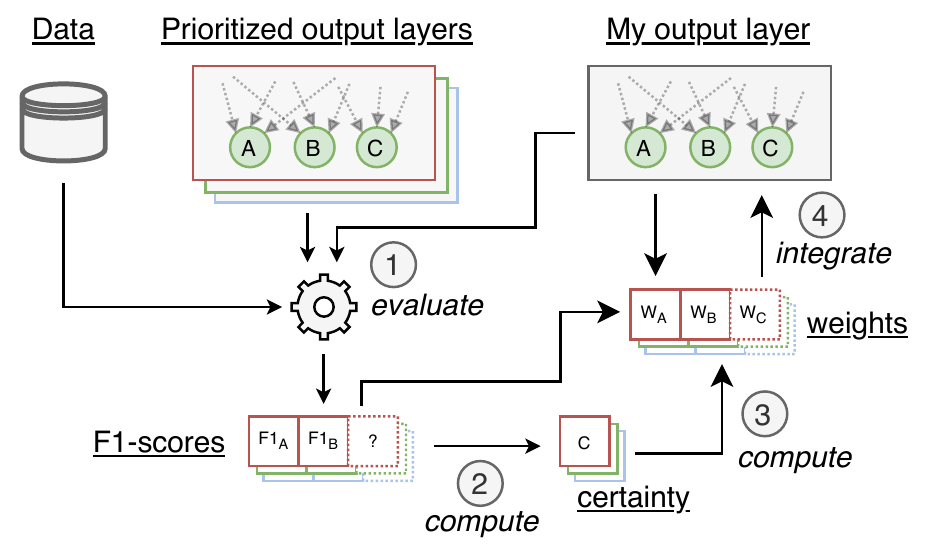}
    \caption{The per-class PBI in Bristle. Dashed boxes refer to values associated with foreign classes.}
    \label{fig:performancebasedintegrator}
\end{figure}

Figure~\ref{fig:performancebasedintegrator} shows how the PBI works, given a model with three classes ($ A $, $ B $ and $ C $).
In summary, the PBI selectively integrates the parameters of classes that perform well to achieve Byzantine-resilience and to handle non-i.i.d. classes properly.
This is different from existing approaches that integrate each received model as a whole instead.
We refer to the parameters in the output layer that are connected to a particular class as  \emph{Class-specific parameters (CSPs)}.
The integration process in Bristle first evaluates the performance of each class of each received model on a test dataset (step~\circled{1}).
Based on this evaluation, we compute a \emph{certainty} score that estimates how benign a received model is (Step~\circled{2}).
Then, we calculate for all CSPs of each model a weight (step~\circled{3}).
Finally, we integrate the received models into our current model, based on the computed weights (step~\circled{4}).

The selective integration of parameters is inspired by the continual learning algorithm CWR*~\cite{RN306} that enables non-i.i.d. learning by initializing the output layer to zero, applying a mean-shift, and replicating the hippocampus-cortex duality by selectively copying and resetting parts of the output layer.
Related work demonstrates that this approach provides excellent performance in regular non-FL environments and enables per-class updates~\cite{RN306}.
Applying continual learning requires that the non-output layers are frozen and equal between all peers.
We already addressed this by using transfer learning (see~Section~\ref{sec:transferlearning}) to train these layers before the first iteration.

We now elaborate on the PBI logic, which pseudocode is given in Listing~\ref{alg:performancebasedintegrator}.
The input for this algorithm is the current model of the peer (\emph{myModel}), and the prioritized models that passed the DBP (\emph{prioritizedModels}).

\begin{algorithm}[t]
	\footnotesize
	\caption{Bristle's performance-based integrator (PBI)}\label{alg:performancebasedintegrator}
	\begin{algorithmic}[1]
		\Procedure{PBI}{\emph{myModel}, \emph{prioritizedModels}}
		\State \emph{F1} $ \gets $ [], \emph{certainty} $ \gets $ [], \emph{disc} $ \gets $ []
		\State \emph{faWg} $ \gets $ [], \emph{foWg} $ \gets $ [] \Comment foreign/familiar class weights
		\For{\emph{m} \textbf{in} \emph{myModel} $ \cup $ \emph{prioritizedModels}} \Comment \textbf{Step 1}
		\For{\emph{c} \textbf{in} \texttt{familiarClasses}(\emph{data})}
		\State \emph{F1}[\emph{m}][\emph{c}] $ \gets $ \texttt{evaluate}(\emph{m}, \emph{c}, \emph{testData})
		\EndFor
		\EndFor
		\State
		
		\For{\emph{m} \textbf{in} \emph{prioritizedModels}}
		\State \emph{best} $ \gets $ \texttt{sorted}(\emph{F1}[\emph{m}])\ [:$ \phi $]
		\State \emph{certainty}[\emph{m}] $ \gets $ \emph{max}(\texttt{avg}(\emph{best}) - \texttt{std}(\emph{best}), 0) \Comment \textbf{Step 2}
		\For{\emph{c} \textbf{in} \texttt{familiarClasses}(\emph{data})} \Comment \textbf{Step 3}
		\If{\emph{F1}[\emph{m}][\emph{c}] $ \ge $ \emph{F1}[\emph{myModel}][\emph{c}]}
		\State \emph{disc}[\emph{m}][\emph{c}] $ \gets $ (|\emph{F1}[\emph{m}][\emph{c}] - \emph{F1}[\emph{myModel}][\emph{c}]| * $ \eta $)$ ^3 $
		\Else
		\State \emph{disc}[\emph{m}][\emph{c}] $ \gets -\infty$
		\EndIf
		\State \emph{faWg}[\emph{m}][\emph{c}] $ \gets $ \texttt{computeFaWg}(\emph{disc}[\emph{m}][\emph{c}], \emph{certainty[\emph{m}]})
		\EndFor
		\State \emph{foWg}[\emph{m}] $ \gets $ \texttt{computeFoWg}(\emph{disc}[\emph{m}], \emph{certainty[\emph{m}]})
		\EndFor
		\State \Comment \textbf{Step 4}
		\State \emph{myModel}.\texttt{integrateFamiliarClasses}(\emph{prioritizedModels}, \emph{faWg})
		\State \emph{myModel}.\texttt{integrateForeignClasses}(\emph{prioritizedModels}, \emph{foWg})
		\EndProcedure
	\end{algorithmic}
\end{algorithm}

\textbf{Step 1 (per-class performance measurements).}
Based on a subset of the local dataset, we first calculate the F1-score for the CSPs of all \emph{familiar classes} of both the peer's own output layer and the prioritized output layers (line 4-8).
We refer to a class as familiar when the peer has a sufficient number of samples, denoted by $ \kappa $, to reliably estimate the performance of that class of a given model.
This threshold is determined by the developer.
With a higher value of $ \kappa $, the PBI can more reliably determine F1-scores, but the number of classes for which insufficient samples are available to evaluate, also known as \emph{foreign classes}, might decrease.
We chose to measure F1-scores instead of the accuracy as a proxy for the performance since the former is more suited for imbalanced datasets~\cite{chicco2020advantages}.
The data subset used by the PBI is never used to train the peer's own model since this would result in an overestimation of the performance of the peer's own model.
Since these measurements depend on the availability of sufficient private data samples, Bristle avoids integrating models from other peers until sufficient private data samples are available to test them reliably.

\textbf{Step 2 (certainty computation).}
Then, we calculate for each prioritized model a \emph{certainty} score, which is a (rough) estimate of the degree to which this model is benign (line 10-12).
This certainty score is determined by subtracting the standard deviation from the average F1-score of the $ \phi $ best-performing familiar classes (line 11-12), thus rewarding high class performance and punishing high variation among the class performance.
We do not consider all classes when computing the certainty score since it is not realistic to assume that all classes of each benign received model perform well in a non-i.i.d. environment.
The larger $ \phi $ is, the more robust the algorithm is against Byzantine attacks, but the less able it is to learn about foreign classes in the case of a non-i.i.d. setting.

\textbf{Step 3 (weight computation).}
Furthermore, we estimate for each familiar class of each prioritized model if integrating its corresponding CSP improves the model's performance by simply checking if the F1-score of that class of the received model exceeds the respective F1-score of the peer's own model (line 14).
We call the extent to which it does the \emph{F1-discrepancy}, and we store this value in a two-dimensional array named \emph{disc}.
If the class-specific F1-score of the model being evaluated exceeds that of our own model, we calculated the F1-discrepancy as in line 15.
Otherwise, we set the F1-discrepancy to $-\infty$ (line 17).
The parameter $\eta$ increases the degree to which high-performing classes are integrated.

The \texttt{computeFaWg} function then computes the weights for the familiar classes for each model, taking the prior computed scores and certainty as input (line 19).
This function computes the following Sigmoid function, where $ w_c $ is the weight of familiar class $ c $, $ s $ is the F1-discrepancy, and $ r $ is the certainty of the model being considered:

\begin{equation}
    w_c = max(0,\frac{\omega_{fa}^1}{1+e^{-\frac{s}{100}}}-\omega_{fa}^2) * r
    \label{eq:sigmoid}
\end{equation}

\begin{figure}[b]
    \includegraphics[width=0.75\linewidth]{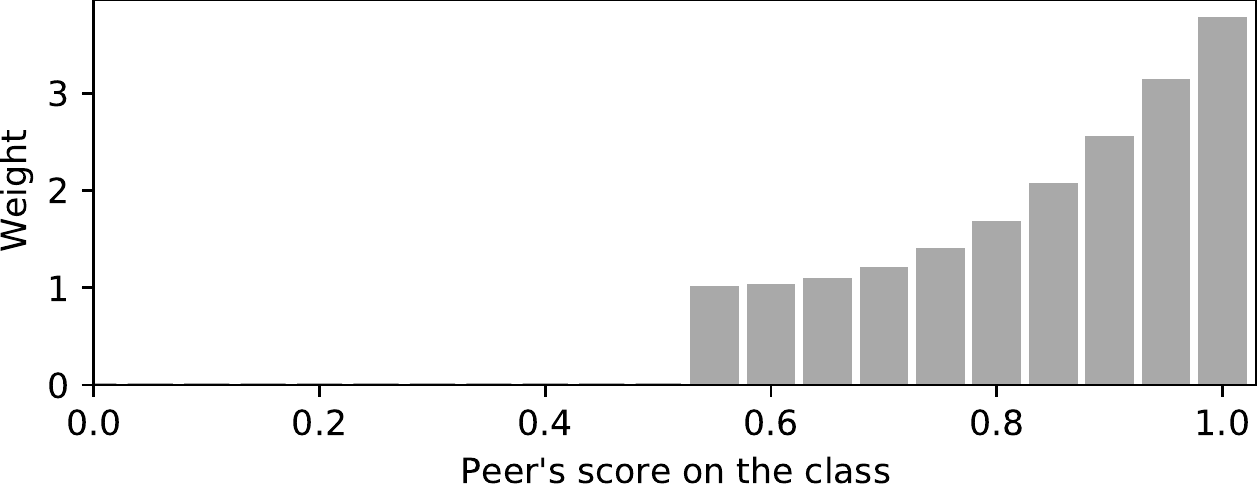}
    \caption{Weight assigned to the CSPs given an F1-score of 0.5 and a certainty score of 1.0.}
    \label{fig:sigmoid}
\end{figure}

This function assigns a weight of one to equally performing classes and an increasingly higher weight to classes that show excellent performance, dependent on the values of $\omega_{fa}^1$ and $\omega_{fa}^2$.
$\omega_{fa}$ represents the boost for above-average performing models and the bounty for below-average performing models. The bigger this discrepancy, the faster the model can catch up with other better-performing models, but the bigger the impact of a malicious model that performs well on familiar classes and bad on foreign classes.
We multiply the outcome of the Sigmoid function by the certainty score calculated earlier because the parameters of a class might be sub-optimal even when it has a perfect F1-score when other classes perform poorly.
Figure~\ref{fig:sigmoid} illustrates how the weight of CSPs is affected by their F1-score on a test dataset, assuming that the peer's current model yields a F1-score of $0.5$ on that class and that the calculated certainty is $1$.

Determining a weight for the CSPs of the foreign classes is challenging because we cannot directly measure their performance.
Instead, we take the sum of the scores of all familiar classes and feed this into the same Sigmoid function as used for the familiar classes (Equation~\ref{eq:sigmoid}), albeit parameterized with separate variables $\omega_{fo}^1$ and $\omega_{fo}^2$.
This is done by the \texttt{computeFoWg} function (line 21).
$\omega_{fo}$ is the extent to which foreign classes are integrated into the model. The higher this value, the better the model can be when the peer wants to use the model to classify formerly foreign classes, but also the higher the impact when a potentially malicious model is integrated.
If the user is uninterested in achieving high performance on foreign classes, $\omega_{fo}^1$ can be set to 0.

\textbf{Step 4 (model integration).}
Finally, we replace the CSPs of familiar and foreign classes with the weighted average of all CSPs by using the calculated weights (line 24-25).

\subsection{Implementation}
\label{sec:implementation}
Bristle is implemented on top of an existing network library that provides support for peer discovery, decentralized overlay creation, and authenticated messaging~\cite{RN279, RN280}.
Since we envision the usage of Bristle mostly in a mobile environment, our middleware is written in the Kotlin programming language (the default language for Android applications).
We envision that Bristle runs as a background service on the end-users' devices to periodically receive or transmit models from and to peers when the device is connected to Wi-Fi.
Peers transmit model updates using the UDP protocol and network messages are compressed with Gzip.
Model training is facilitated by the DeepLearning4j library (version 1.0.0-beta7) to enjoy compatibility with a wide range of advanced ML algorithms.
We have published the Bristle source code and developer documentation in a GitHub repository.\footnote{Source code can be requested through the Program Committee (to comply with the double-blind review requirements).}

\textbf{Using Bristle.}
Developers can leverage the Bristle middleware by feeding the peer's current model and all received models into Bristle's GAR after each iteration and subsequently replacing the peer's model with the result.
The moment a model is updated can be decided by the developer, for example, model training can take place when the device is charging to minimize the impact on end users.
The peers should be able to selectively receive and integrate only models relevant to the current ML application.
To this end, Bristle uses the functionality to form communication groups as provided by its network library.
Deciding on the variables listed under Section~\ref{sec:experimental_setup} is a key part of machine learning.
We recommend the developer to use heuristics, optimize them on a related dataset, or use A/B-testing to find the best values to obtain the desired accuracy on the models.

\section{Experimental Setup}
\label{sec:experimental_setup}
We now describe our experimental setup, datasets, and parameters.

\textbf{Testbed.}
All experiments are run on an HPE DL385 Gen10 server.
This server is equipped with 128 AMD EPYC 7452 CPUs, has 512 GB of DDR4 memory, and runs Debian 10.

\textbf{Datasets.}
In line with related research~\cite{RN79,RN6,RN91}, we consider an image classification application that applies Convolutional Neural Networks (CNNs) to classify images.
We use the popular MNIST~\cite{RN1012} dataset, consisting of 60,000 gray-scale training images and 10,000 test images of 28x28 pixels representing handwritten digits.
To achieve better performance, we standardize the dataset by applying Z-score normalization such that the features are re-scaled to a normal distribution with $\mu$ = 0 and $\sigma$ = 1.
We pre-train - until convergence - a neural network with the same configuration on the EMNIST-Letters~ \cite{cohen2017emnist} dataset, which features resemble MNIST, but contains characters instead of digits.
We then include the resulting neural network in the Bristle software.

\begin{table}[t]
\begin{tabular}{|l|l|} \hline
\textbf{Experiment parameter} & \textbf{Default value} \\ \hline

\emph{Environment} & \\
Peers ($ n $) & 10 \\
Connection ratio & 100\% \\
Fraction Byzantine peers & 50\% \\ \hline

\emph{Machine learning} & \\
Mini-batch size ($ b $) & 5 \\
Learning rate ($ \lambda $) & 0.001 \\
L2-value & 0.005 \\
Max. iterations & 300 \\ \hline

\emph{Bristle} & \\
Exploration-exploitation rate ($\alpha$) & 0.4 \\
Max. PBI input size ($ \beta $) & 30 \\
\#familiar class selection size ($\phi$) & 3 \\
\#test samples per class ($\kappa$) & 10 \\
$ \eta $ & 10 \\
$\omega_{fa}^1$, $\omega_{fo}^1$ & 10 \\
$\omega_{fa}^2$, $\omega_{fo}^2$ & 4 \\  \hline
\end{tabular}
\caption{Default parameters and values used during the experiments.}
\label{tab:experiment_parameters}
\end{table}

\textbf{Experiment parameters.}
We list all default parameters used during our experiments in Table~\ref{tab:experiment_parameters}.
An exploration-exploitation ratio $ \alpha $ of 0.4 slightly prioritizes model with lower distance.

\textbf{Non-i.d.d. data. }
To evaluate Bristle with non-i.d.d. data, we first sort the data per class, divide the data into equally-sized shards, and then assign to every peer several shards, which is also the approach taken by related work~\cite{RN1002, RN257, RN271}.
We note that the more shards we assign to every peer, the better every peer can recognize and defend against Byzantine attacks (see Section~\ref{sec:performancebasedintegrator}), but the less non-i.i.d. the classes are.
We opt to assign - unless specified otherwise - without loss of generality to every peer four shards which cover 40\% of the classes to balance between the ability to learn non-i.i.d. classes and to recognize Byzantine attacks.

\textbf{Model training.}
We train the dataset on the same CNN architecture used by McMahan et al. \cite{RN6}, except that we use Leaky ReLu instead of the regular ReLu as the activation function for the hidden layers since the former seems to give slightly better performance (specifically,  suffers less from the vanishing gradients problem).
The model consists of twice a convolutional layer (kernel size = 5, stride = 1, padding = 0) followed by a max pooling layer (kernel size = 2, stride = 2, padding = 0), and finally the output layer (with 800 hidden nodes).
For the output function, we use the softmax function, and as the loss function, we use negative log-likelihood.
To train the model, we use the Adam optimizer, parameterized as shown in Table~\ref{tab:experiment_parameters}.
Since these values significantly impact the model's performance, we used a grid search with typically 7-9 values for each parameter and ensured that the optimal values were approximately in the middle.
Although the baselines were originally not developed to be used in combination with transfer learning, we decided to use transfer learning for all baselines for all experiments anyway since the performance increase is so significant that otherwise any comparison with Bristle would be meaningless (also see Section~\ref{sec:regvstrans}).

\textbf{Baselines.}
To compare Bristle with existing methods, we implemented five other GARs commonly used in FL:
\begin{enumerate}
	\item \emph{FedAvg}~\cite{RN6} aggregates all models by coordinate-wise averaging of parameters. It is commonly used as a baseline to compare the performance of FL systems.
	\item \emph{Median} \cite{RN11} aggregates all models by taking the coordinate-wise mean of parameters. As demonstrated in the literature, it is already a particularly effective Byzantine-resilient GAR~\cite{RN196}.
	\item \emph{Krum} \cite{RN12} integrates the model that most closely resembles (in terms of Euclidean distance) all other models as the new global model. Even if the selected model is malicious, in theory, the performance should not degrade too much as it is close to all other models.
	\item \emph{BRIDGE} \cite{RN131} is specifically designed for Byzantine-resilient model aggregation in decentralized settings. It cyclically updates every coordinate one by one and subsequently applies trimmed-mean screening to obtain the final coordinate for each dimension.
	\item \emph{MOZI} \cite{RN132} uses a combination of a fast distance-based and accurate performance-based filter to aggregate model updates in a Byzantine-resilient manner.
\end{enumerate}
We initialized Krum and BRIDGE, which are dependent on a-priori knowledge of the number of attackers, with $ b = 4 $ (the maximum number of attackers) and Mozi with $\rho = 0.5$ (the ratio of benign to Byzantine peers).

\begin{figure}[b]
    \centering
    \begin{subfigure}{0.495\linewidth}
        \centering
        \includegraphics[width=\textwidth]{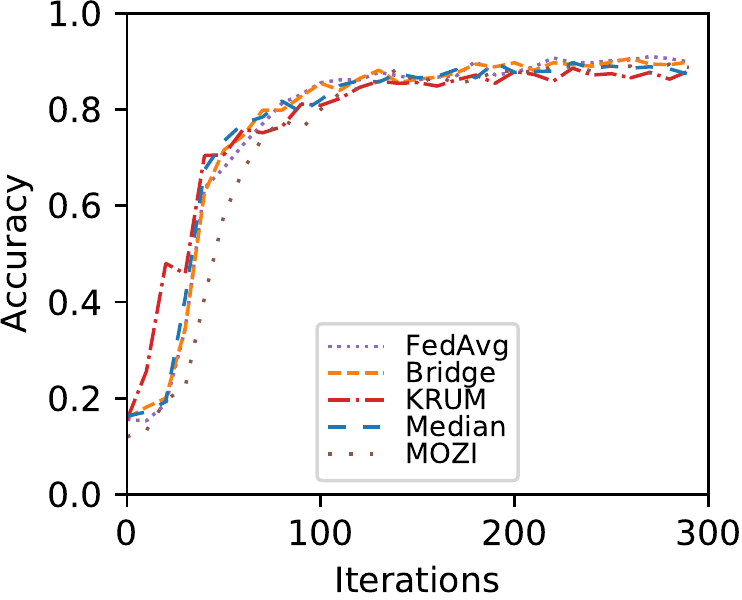}
        \caption{Regular learning}
        \label{fig:exp_regular_learning}
    \end{subfigure}
    \hfill
    \begin{subfigure}{0.495\linewidth}
        \centering
        \includegraphics[width=\textwidth]{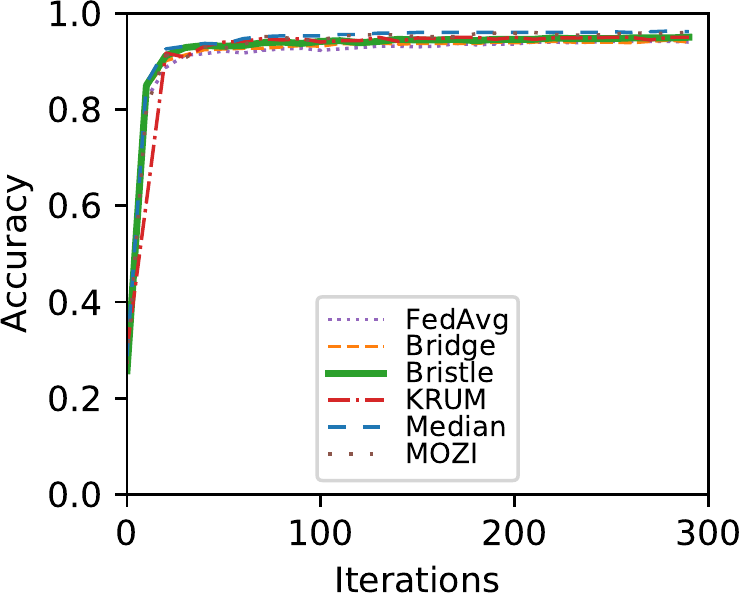}
        \caption{Transfer learning}
        \label{fig:exp_transfer_learning}
    \end{subfigure}
    \caption{The accuracy of the model while training, for regular and transfer learning.}
    \label{fig:exp_regular_transfer_learning}
\end{figure}

\begin{figure*}[t]
    \centering
    \begin{subfigure}[b]{0.245\textwidth}
        \centering
        \includegraphics[width=\textwidth]{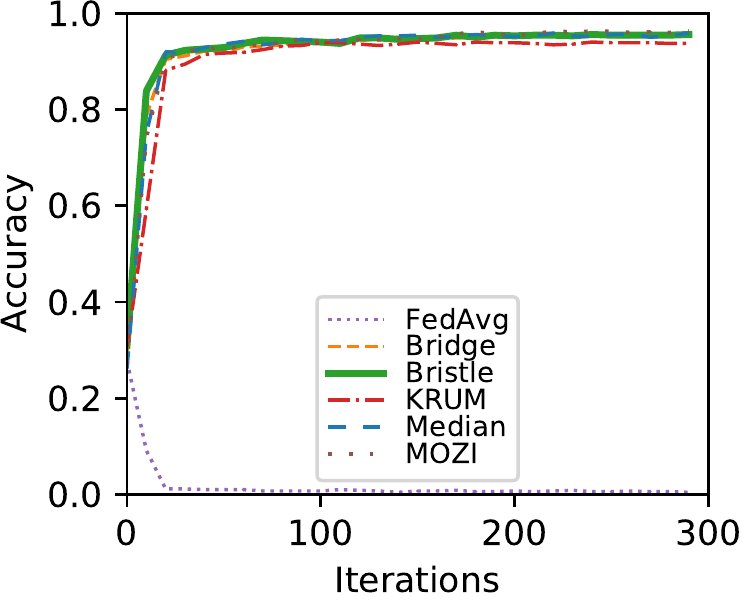}
        \caption{Label-flip attack}
        \label{fig:exp_all_label_flip_attack}
    \end{subfigure}
    \hfill
    \begin{subfigure}[b]{0.245\textwidth}
        \centering
        \includegraphics[width=\textwidth]{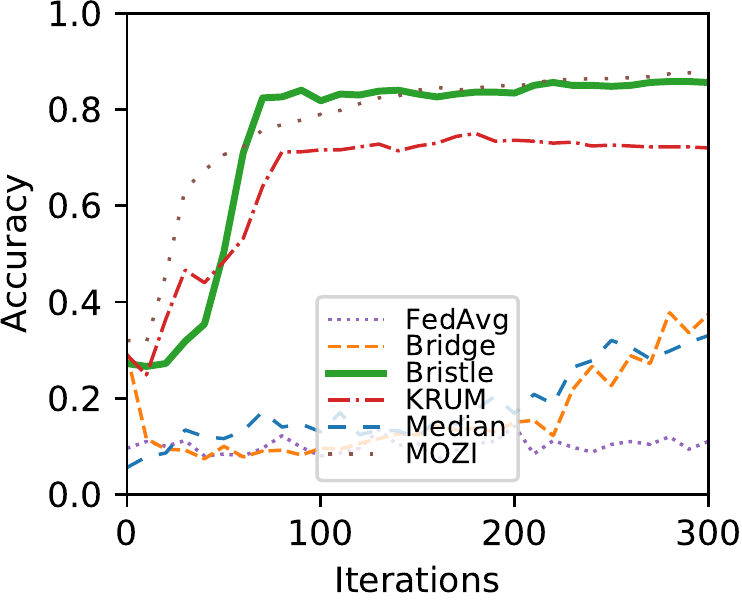}
        \caption{Additive noise attack}
        \label{fig:exp_additive_noise_attack}
    \end{subfigure}
    \hfill
    \begin{subfigure}[b]{0.245\textwidth}
        \centering
        \includegraphics[width=\textwidth]{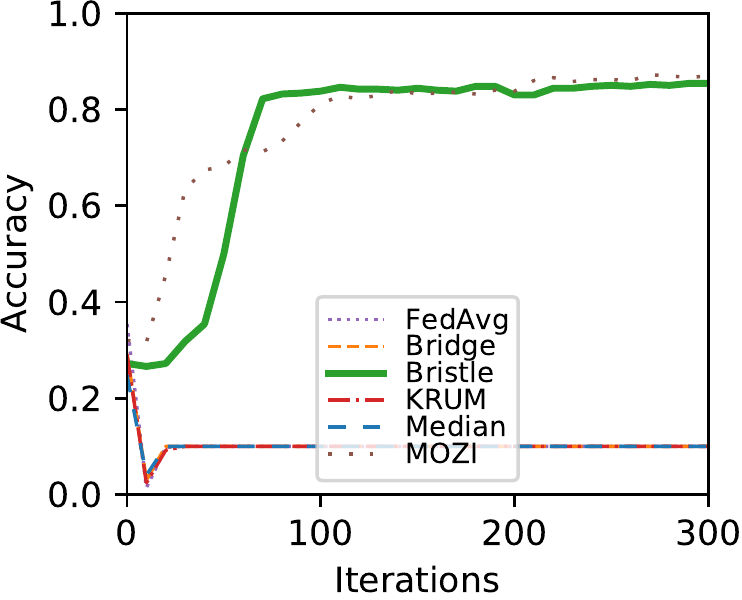}
        \caption{Krum attack}
        \label{fig:exp_krum_attack}
    \end{subfigure}
    \hfill
    \begin{subfigure}[b]{0.245\textwidth}
        \centering
        \includegraphics[width=\textwidth]{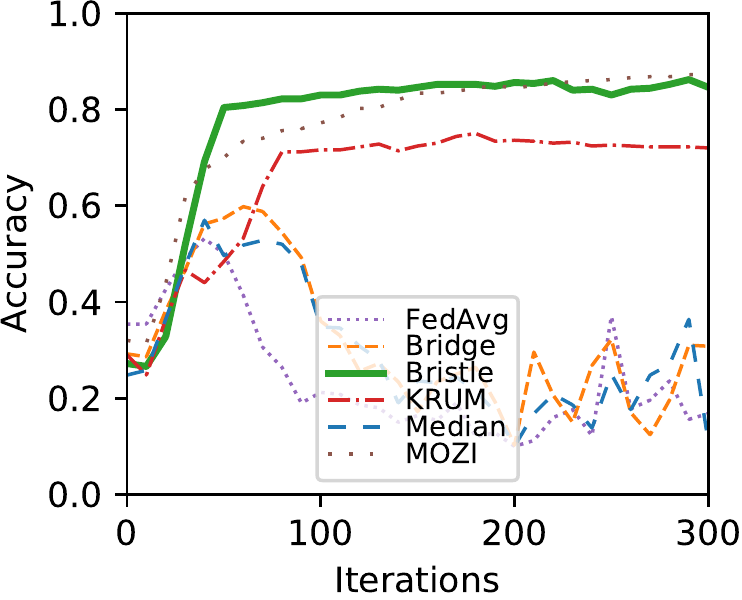}
        \caption{Trimmed Mean attack}
        \label{fig:exp_trimmed_mean_attack}
    \end{subfigure}
    \caption{The resilience of Bristle and other GARs against various Byzantine attacks, with 50\% of all peers being malicious and i.i.d. data.}
    \label{fig:exp_byzantine_resilience}
\end{figure*}

\textbf{Byzantine attacks. }
We also evaluate Bristle under the following four Byzantine attacks, which are commonly considered in the domain.
\begin{enumerate}
    \item The \emph{Label-flip attack}~\cite{RN21} assigns an incorrect label to each input~\cite{RN21, RN40, RN311}. We implement this attack by numbering all labels and reassigning each sample with label $x$ to label $(x + 1)\ \%\ |x|$.
    \item The \emph{Additive noise attack}~\cite{RN184} adds some noise to the parameters of outgoing models. When the noise has a larger variance, it can indeed prevent convergence but also makes the noise attack easier to detect~\cite{RN184}. Centering the noise around a value slightly different from 0 allows the attack to prevent convergence despite low variance, but since the mean of benign updates is always centered around 0, this attack can be easily detected. We consider a variant where each half of the parameters are set to noise centered around a value just below and above 0, respectively.
    \item The \emph{Krum attack}~\cite{RN17} specifically targets the Krum aggregation rule. It is an effective, state-of-the-art attack by iteratively sending attack vectors that will be accepted by Krum whilst inflicting maximum damage to the peer’s model.
    \item The \emph{Trimmed Mean attack}~\cite{RN17} targets the trimmed mean GAR (Bridge in our experiments). It determines the gradient direction for each parameter of the model and then creates an attack vector that points exactly in the opposite direction, scaled per parameter depending on the values of the other benign peers.
\end{enumerate}

\section{Experimental Evaluation}\label{sec:results}
We now evaluate the performance of Bristle.
Our evaluation answers the following questions: \emph{(1) what is the achieved training speedup when applying transfer learning to DFL? (2) How does Bristle perform in the presence of Byzantine attackers in terms of model accuracy? (3) How does Bristle perform when classes are not uniformly distributed over peers (non-i.i.d.) in terms of model accuracy? (4) How does Bristle perform in an environment with both Byzantine attackers and non-i.i.d. classes? And (5) What are the communication and computational costs of Bristle?}

\begin{figure}[t]
    \centering
    \begin{subfigure}[b]{0.495\linewidth}
        \centering
        \includegraphics[width=\textwidth]{Figure_1.1_-_transfer_-_MNIST_-_10_nodes.pdf}
        \caption{All peers own samples of all classes}
        \label{fig:exp_regular}
    \end{subfigure}
    \hfill
    \begin{subfigure}[b]{0.495\linewidth}
        \centering
        \includegraphics[width=\textwidth]{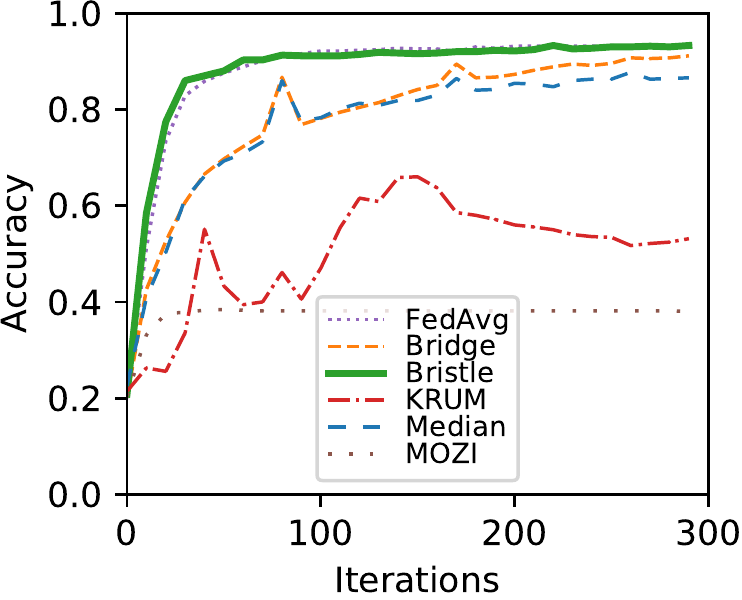}
        \caption{Each peer owns samples of 40\% of the classes}
        \label{fig:exp_transfer}
    \end{subfigure}
    \caption{The performance of Bristle and other GARs when the data is i.i.d. and non-i.i.d.}
    \label{fig:exp_non_iid}
\end{figure}

\subsection{Performance Gains of Transfer Learning}\label{sec:regvstrans}
We first evaluate the performance gains of transfer learning when all peers are honest, and the data is i.i.d.
For each experiment, we measure after every 10 iterations the performance of all peers and then take the average accuracy, defined as $\frac{\#correct\ predictions}{\#test\ samples}$.
Figure~\ref{fig:exp_regular_transfer_learning} shows the evolution of model accuracy for the baseline GARs and Bristle, both without transfer learning (Figure~\ref{fig:exp_regular_learning}) and with transfer learning (Figure~\ref{fig:exp_transfer_learning}).
We note that Bristle requires a pre-trained model and therefore is not included in Figure~\ref{fig:exp_regular_learning}.
Figure~\ref{fig:exp_regular_learning} shows that the evaluated GARs perform quite similarly, although Mozi has a slower convergence rate since it assigns lower weights to benign models that have not been trained sufficiently yet to perform well.
Transfer learning dramatically improves both the convergence rate and the maximum accuracy after 300 iterations of all GARs, which also supports literature~\cite{joy2019flexible}.
More specifically, whereas reaching 70\% accuracy takes on average 55 iterations with regular training, it takes merely four iterations with transfer learning, a reduction of 93\%.
Reaching 90\% accuracy takes on average 180 iterations with regular learning, whereas it only takes 30 iterations with transfer learning, a reduction of 83\%.
This reduction in iterations to reach the desired level of model accuracy can  significantly reduce the load on the network.

\begin{figure*}[t]
    \centering
    \begin{subfigure}[b]{0.245\textwidth}
        \centering
        \includegraphics[width=\textwidth]{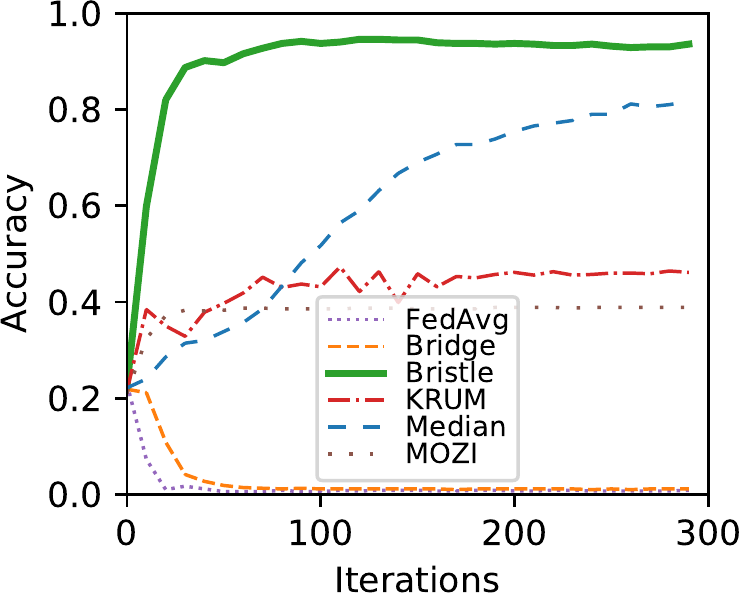}
        \caption{Label-flip attack}
        \label{fig:exp_all_label_flip_attack_noniid}
    \end{subfigure}
    \hfill
    \begin{subfigure}[b]{0.245\textwidth}
        \centering
        \includegraphics[width=\textwidth]{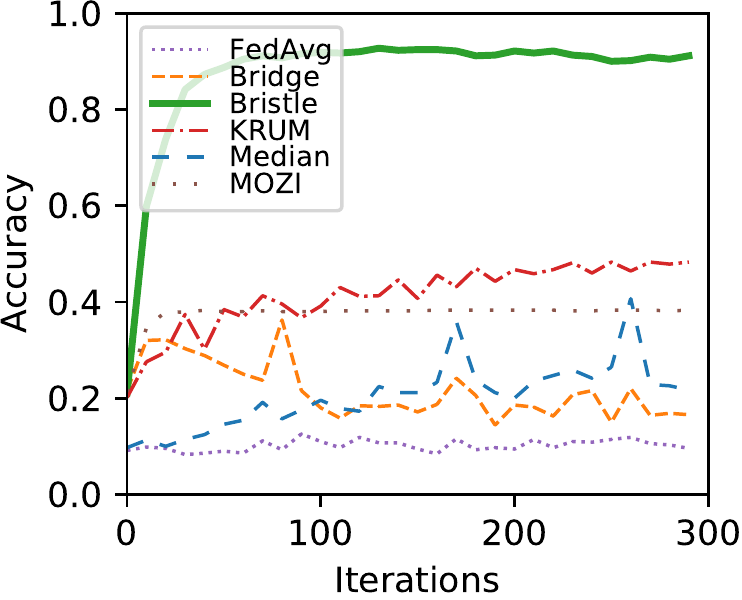}
        \caption{Additive noise attack}
        \label{fig:exp_additive_noise_attack_noniid}
    \end{subfigure}
    \hfill
    \begin{subfigure}[b]{0.245\textwidth}
        \centering
        \includegraphics[width=\textwidth]{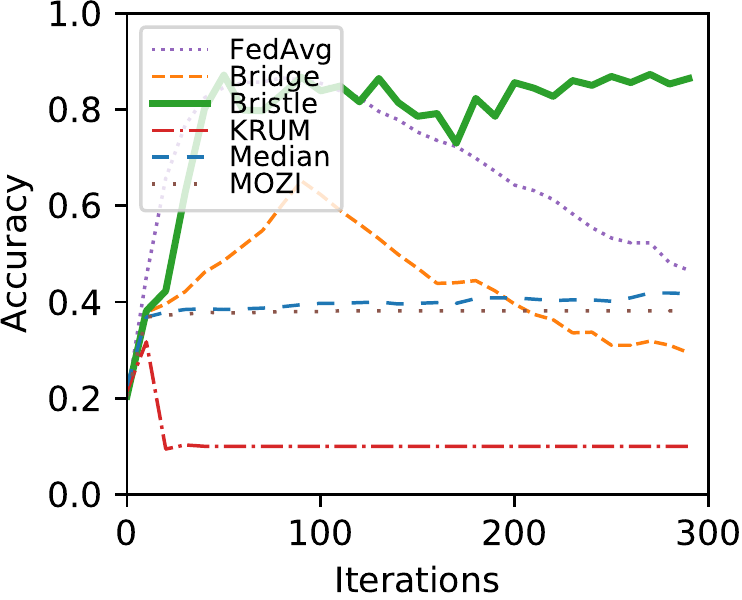}
        \caption{Krum attack}
        \label{fig:exp_krum_attack_noniid}
    \end{subfigure}
    \hfill
    \begin{subfigure}[b]{0.245\textwidth}
        \centering
        \includegraphics[width=\textwidth]{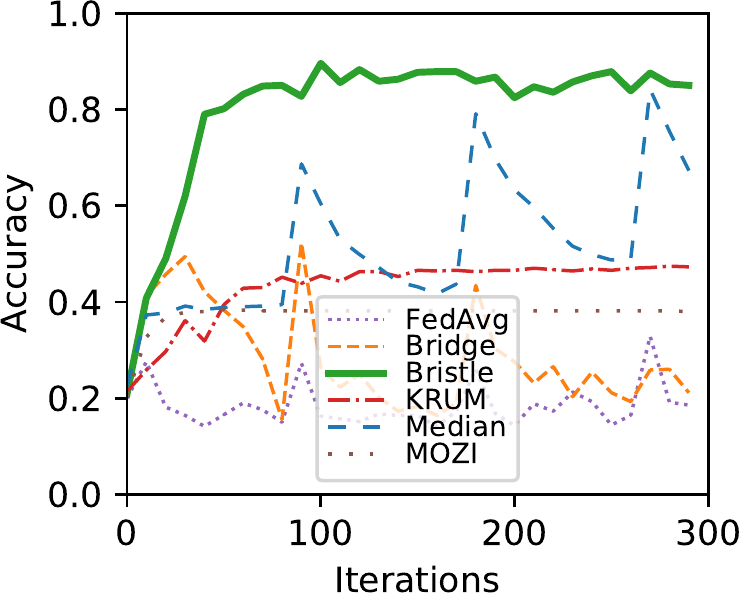}
        \caption{Trimmed Mean attack}
        \label{fig:exp_trimmed_mean_attack_noniid}
    \end{subfigure}
    \caption{The resilience of Bristle and other GARs against various Byzantine attacks when the data is non-i.i.d.}
    \label{fig:exp_byzantine_resilience_noniid}
\end{figure*}

\begin{figure*}[t]
    \centering
    \begin{subfigure}[b]{0.245\textwidth}
        \centering
        \includegraphics[width=\textwidth]{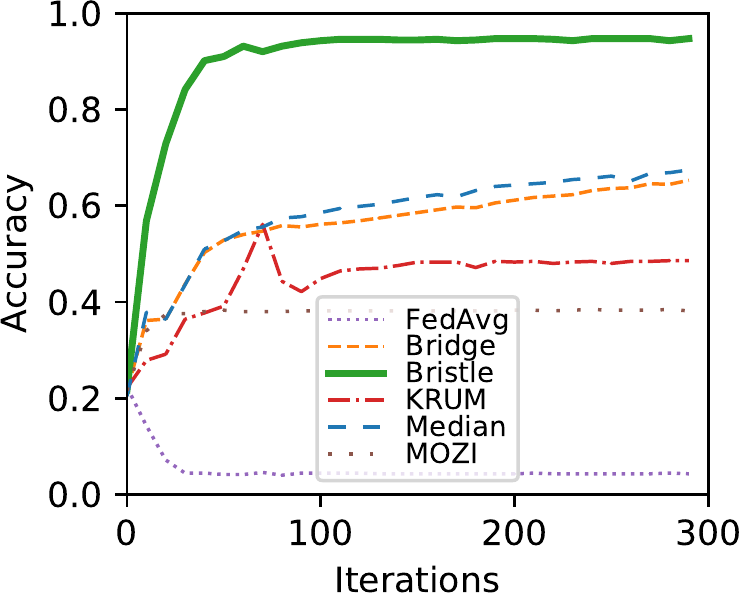}
        \caption{10\% of the peers is malicious}
        \label{fig:exp_numattackers_10}
    \end{subfigure}
    \hfill
    \begin{subfigure}[b]{0.245\textwidth}
        \centering
        \includegraphics[width=\textwidth]{Figure_4.3_-_transfer_-_MNIST_-_10_nodes_non-i.i.d._40p_+_5_all-label-flip_attackers_30p.pdf}
        \caption{30\% of the peers is malicious}
        \label{fig:exp_numattackers_30}
    \end{subfigure}
    \hfill
    \centering
    \begin{subfigure}[b]{0.245\textwidth}
        \centering
        \includegraphics[width=\textwidth]{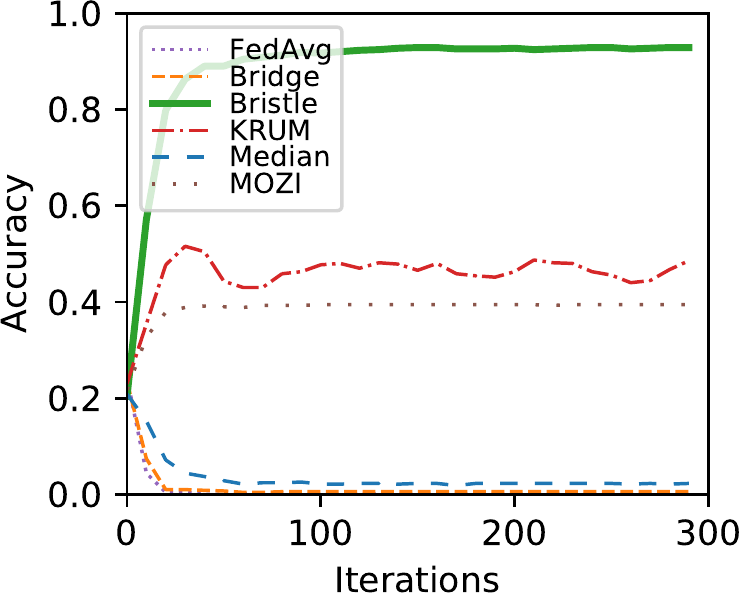}
        \caption{50\% of the peers is malicious}
        \label{fig:exp_numattackers_50}
    \end{subfigure}
    \hfill
    \begin{subfigure}[b]{0.245\textwidth}
        \centering
        \includegraphics[width=\textwidth]{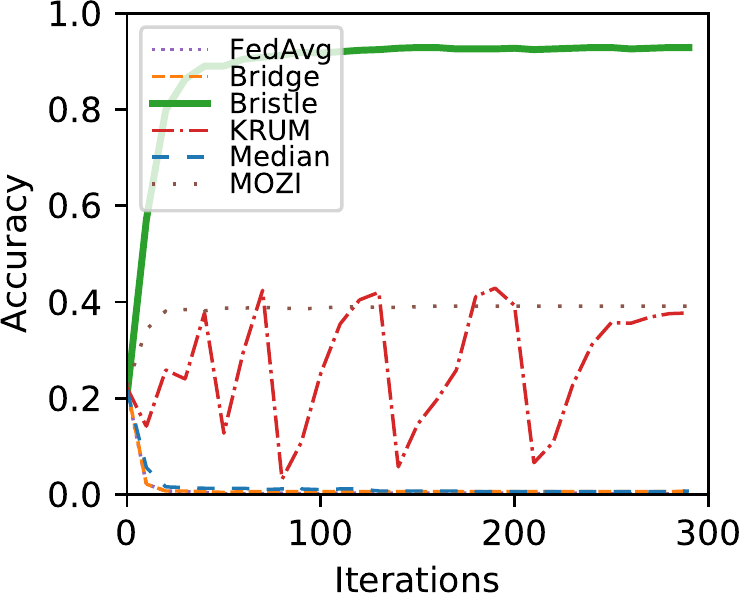}
        \caption{70\% of the peers is malicious}
        \label{fig:exp_numattackers_70}
    \end{subfigure}
    \caption{The resilience of Bristle and other GARs under a label-flip-attack, with a varying percentage of all peers being malicious, where the data is non-i.i.d.}
    \label{fig:exp_numattackers_comparison}
\end{figure*}

\subsection{Byzantine Attacks}\label{sec:exp_byzantine_resilience}
We now evaluate the Byzantine-resilience of Bristle under different attacks when data is i.i.d.
Figure~\ref{fig:exp_byzantine_resilience} shows the effect of different attacks.
Bristle withstands all evaluated Byzantine attacks and quickly achieves high model accuracy.
Figure~\ref{fig:exp_all_label_flip_attack} shows that the label-flip attack prevents model convergence when using the FedAvg GAR, but is, despite its relatively small influence on the model’s parameters, successfully mitigated by all other GARs.
Figure~\ref{fig:exp_additive_noise_attack} shows that the FedAvg and Median GARs are susceptible to the additive noise attack.
The Krum attack, shown in Figure~\ref{fig:exp_krum_attack}, is clearly very effective against Krum but has only a minor effect on the convergence rate of the other GARs.
The Trimmed Mean attack, see Figure~\ref{fig:exp_trimmed_mean_attack}, is relatively ineffective against any Byzantine-resilient GAR. This is because all benign models are very close to each other in this scenario, making it hard for this attack to steer the model in another direction without clearly being an outlier.

\subsection{Non-i.i.d. Classes}\label{sec:noniiddata}
We now consider the scenario where the classes are non-i.i.d., see Figure~\ref{fig:exp_non_iid}.
In this experiment, there are 10 peers, each of which has access to two random samples of four consecutive classes.
Thus, each peer $i$ has access to classes $ \{i, (i+1) \% 10, (i+2) \% 10, (i+3) \% 10\} $.
When we compare Figure~\ref{fig:exp_transfer} with Figure~\ref{fig:exp_regular}, it is clear that Krum and Mozi fail to achieve desired model accuracy with non-i.i.d. classes, and we also observe that Bridge and Median converge significantly slower.
FedAvg and Bristle show excellent performance and achieve 90\% model accuracy in 65 and 55 iterations, respectively.
This is because these methods (eventually) combine the information learned by every peer, while the other methods disregard a part of the received models under the incorrect assumption that those are Byzantine.

\subsection{Combining Byzantine Attacks and Non-i.i.d. Classes}
We now evaluate Bristle and other GARs in an environment containing both Byzantine attackers and where the classes are non-i.i.d. (see Figure~\ref{fig:exp_byzantine_resilience_noniid}).
We use the label-flip attack in all experiments because this one is effective (see Figure~\ref{fig:exp_all_label_flip_attack_noniid}) and very popular in related work~\cite{RN133, RN310, RN17}.
Except for Bristle, all baselines fare poorly in the Byzantine experiments with non-i.i.d. classes.
Specifically, we observe that FedAvg is unable to defend against Byzantine attacks, although it achieves an accuracy of 87\% in the first few iterations of the Krum attack (see Figure~\ref{fig:exp_krum_attack_noniid}).
Although Bridge performs well in the experiments presented in Section~\ref{sec:exp_byzantine_resilience}, it performs poorly under the threat of Byzantine attacks when the classes are non-i.i.d.
Krum is, as expected, unable to defend against the Krum attack but performs under the threat of the other attacks roughly on par with Mozi.
Mozi shows the same accuracy as in the benign non-i.i.d. experiments (see Figure~\ref{fig:exp_byzantine_resilience}) because it successfully defends against Byzantine attacks but is also maxed out on the poor maximum accuracy that it can attain on the node's own dataset.
The performance of the Median baseline varies significantly depending on the attack and may be quite inconsistent (as illustrated in Figure~\ref{fig:exp_additive_noise_attack_noniid} and Figure~\ref{fig:exp_trimmed_mean_attack_noniid}).
An interesting finding is that the Trimmed Mean attack is relatively ineffective in the i.i.d. experiment (see Figure~\ref{fig:exp_trimmed_mean_attack}) but impacts the performance of the same baselines significantly in the non-i.i.d. experiment (see Figure~\ref{fig:exp_trimmed_mean_attack_noniid}).
This results from the greater distance between the benign models in the non-i.i.d. experiment, giving the attack more leeway to steer the model in a different direction without being considered an outlier.
Since the PBI assigns a weight of 0 to the malicious CSPs, Bristle consistently outperforms all other GARs and defends well against all evaluated Byzantine attacks.

\begin{figure}[t]
    \centering
    \begin{subfigure}[b]{0.495\linewidth}
        \centering
        \includegraphics[width=\textwidth]{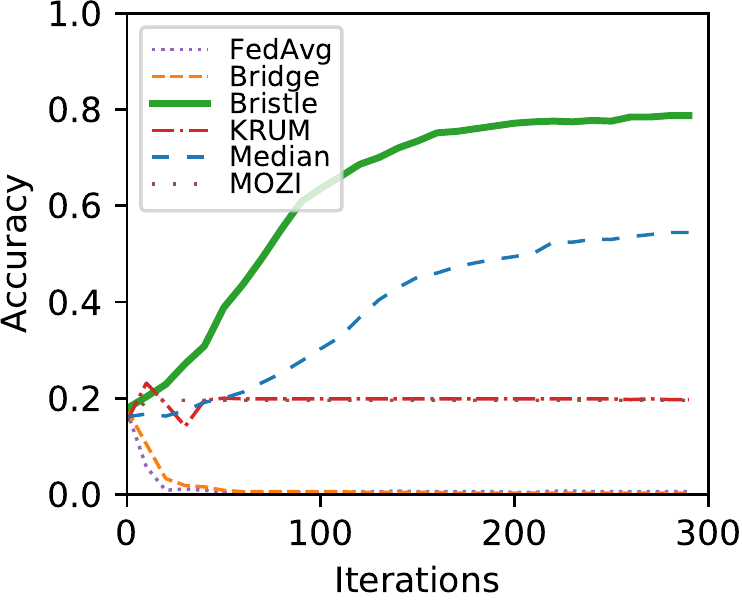}
        \caption{Each peer has data of 20\% of the classes}
        \label{fig:exp_noniid_20}
    \end{subfigure}
    \hfill
    \begin{subfigure}[b]{0.495\linewidth}
        \centering
        \includegraphics[width=\textwidth]{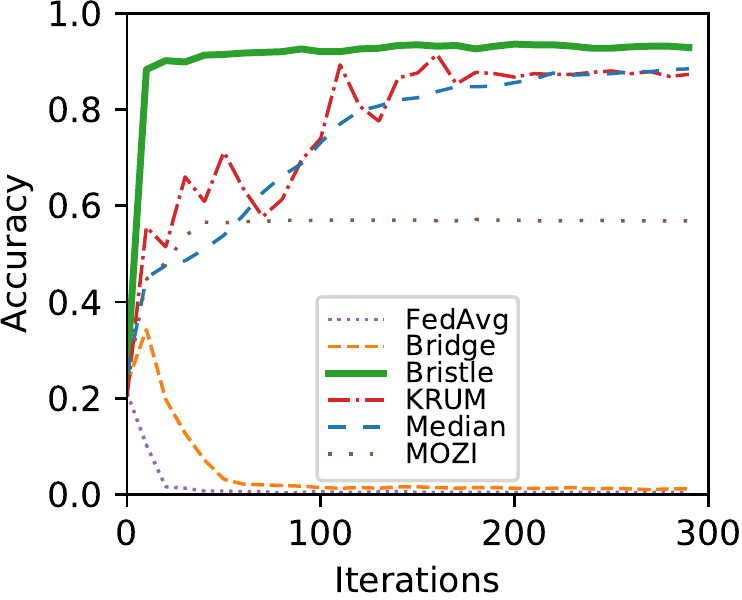}
        \caption{Each peer has data of 60\% of the classes}
        \label{fig:exp_noniid_60}
    \end{subfigure}
    \caption{The resilience of Bristle and other GARs under a label-flip-attack, with 30\% of all peers being malicious, where the classes are to a varying extent non-i.i.d.}
    \label{fig:exp_noniid_comparison}
\end{figure}
\begin{figure}[t]
    \centering
    \begin{subfigure}[b]{0.495\linewidth}
        \centering
        \includegraphics[width=\textwidth]{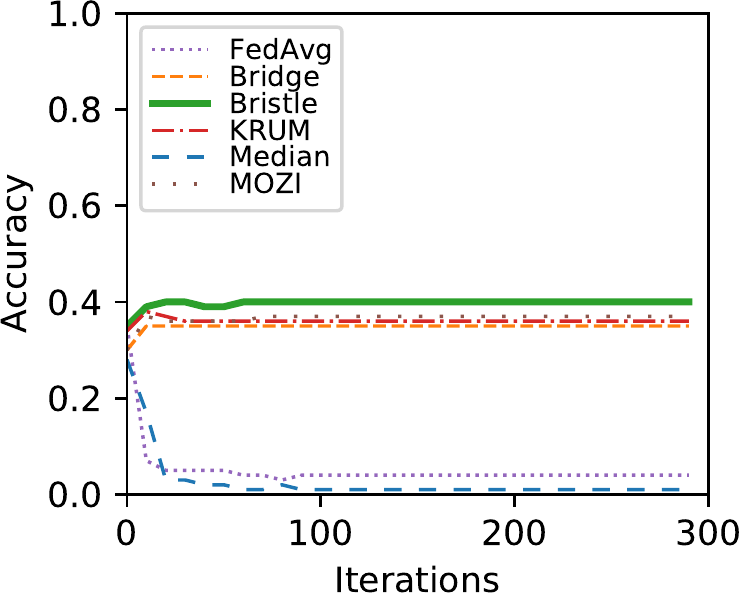}
        \caption{Each peer is connected to 2\% of the other peers}
        \label{fig:conrat2}
    \end{subfigure}
    \hfill
    \begin{subfigure}[b]{0.495\linewidth}
        \centering
        \includegraphics[width=\textwidth]{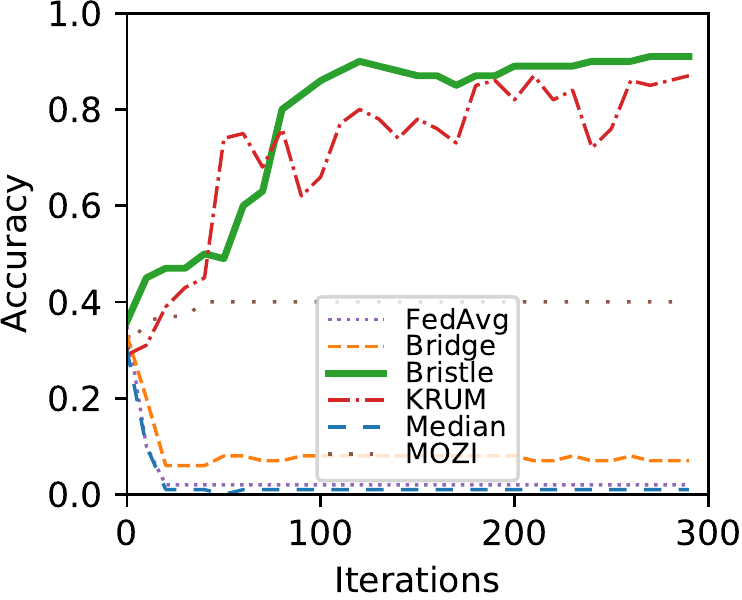}
        \caption{Each peer is connected to 5\% of the other peers}
        \label{fig:conrat5}
    \end{subfigure}
    \caption{The resilience of Bristle and other GARs under a label-flip-attack, with a varying percentage of the peers connected, where the classes are non-i.i.d.}
    \label{fig:exp_connectionrate_comparison}
\end{figure}

\textbf{Varying the Number of Byzantine Attackers.}
We now vary the number of Byzantine attackers, see Figure~\ref{fig:exp_numattackers_comparison}.
Figure~\ref{fig:exp_numattackers_10} shows that even with 10\% Byzantine attackers, the performance of the baselines already degrades significantly.
Bristle manages to maintain a quick increase in accuracy for all considered attack scenarios.
With 10\% and 30\% Byzantine attackers, Mozi manages to keep a stable performance but is limited to predicting only the peer's familiar classes correctly.
Figure~\ref{fig:exp_numattackers_70} shows that the performance of Krum is inconsistent and that only Bristle achieves a consistent high performance.
Bristle's excellent performance results from the fact that (a) in contrast to Mozi, Bristle evaluates and integrates the parameters per class instead of per model, and (b) in contrast to the other GARs, Bristle uses performance evaluations instead of just distance comparisons.

\textbf{Varying the Degree of Non-i.i.d.-ness.}
We now compare the performance of the GARs in three situations where the classes are to a varying extent non-i.i.d. (see Figure~\ref{fig:exp_noniid_comparison}).
FedAvg and Bridge are, regardless of the degree of non-i.i.d.-ness, unable to provide any resilience to the label-flip attack.
Mozi defends, similarly to the experiments in Figure~\ref{fig:exp_byzantine_resilience_noniid}, perfectly well against the label-flip attack but is limited to the maximum accuracy that can be obtained by training on its own dataset.
The performance of Krum increases with the number of classes owned by the peer.
The Median rule performs relatively well, even when the classes are highly non-i.i.d.
Bristle, however, clearly performs best compared to all baselines, although its performance decreases when the peers have only 20\% of the data (see Figure~\ref{fig:exp_noniid_20}).

\textbf{Varying the Connection Ratio.}
In larger networks, it is unrealistic to assume all-to-all dissemination of model updates.
To test the impact of the connection rate on the convergence rate more accurately, we set up an environment with 100 peers where the classes are again non-i.i.d. similarly to the previous experiments.
We setup two experiments, connect each peer to a random subset of 2\% and 5\% of the other peers respectively, add an equal number of label-flip attackers that are connected to each benign peer, and measure the average accuracy over time.
From Figure~\ref{fig:conrat2}, we observe that when the connection rate is only 2\%, FedAvg and Median perform very poorly, but also the other GARs are unable to achieve satisfactory accuracy.
When the connection rate increases to 5\% in Figure~\ref{fig:conrat5}, Bristle and Krum perform quite well in contrast to the other baselines.
Thus, it seems that in a setting with 100 peers, a connection rate of only 5\% is already almost enough to reach the maximum accuracy.
The number of iterations required to obtain this accuracy is relatively high compared to a connection rate of 100\% (see Figure~\ref{fig:exp_all_label_flip_attack_noniid}).

\begin{figure}[b]
    \centering
    \includegraphics[width=0.75\linewidth]{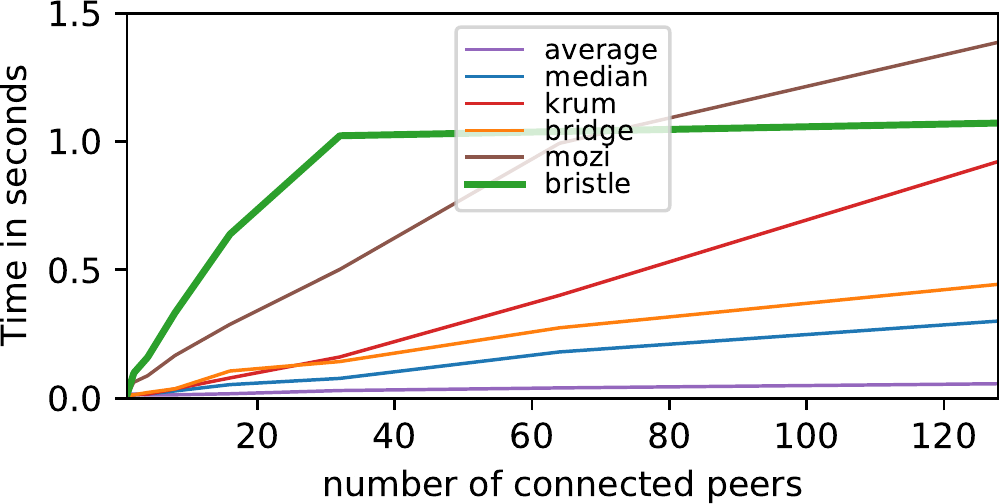}
    \caption{Average time to complete a single training iteration on the non-i.i.d. label-flip experiment with a varying number of connected peers.}
    \label{fig:exp_speed}
\end{figure}

\subsection{Bristle Efficiency and Scalability}
Figure~\ref{fig:exp_speed} shows for each GAR the average time it takes for a peer to finish a single iteration depending on the number of connected peers.
For evaluation purposes we assume that each peer receives from every other peer exactly one model at every iteration.
The running time of Average is clearly the fastest, followed closely by Median, Krum, and Bridge.
The fast execution of these GARs results from the usage of relatively inexpensive operations, such as calculating the distance between models and sorting the parameters of each dimension.
Mozi is relatively slow because it has to evaluate the accuracy of each model.
Bristle is initially by far the slowest GAR caused by the performance-based integrator that evaluates the performance not just for each model but for each familiar class of each model.
However, when the number of connected peers exceeds $\beta$ (see Section~\ref{sec:distancebasedprioritizer}), adding more peers has a negligible impact on the speed of Bristle because the distance-based prioritizer is efficient in reducing the number of models to $\beta$.

All baselines consume an equal amount of network traffic (roughly 3 MB per iteration per peer for our neural network), but because Bristle transmits only the output layer (roughly 300 KB), the bandwidth requirements of Bristle are reduced by 90\%.

\section{Related work}
In the last five years, a considerable amount of literature has been devoted to creating more effective FL systems.
Most work focuses on achieving Byzantine-resilience but also introduces several methods to reduce the bandwidth requirements and to improve learning on non-i.i.d. classes.

\textbf{Byzantine-resilience. }
Several well-known Byzantine-resilient GARs are Coordinate-wise Median (CM) \cite{RN310} which takes the median across all models for each parameter, Krum \cite{RN12} which selects the model that most closely resembles all other models, and Bulyan \cite{RN79} which iteratively applies Krum followed by a variant of CM.
However, these GARs use distances as a proxy for benignness, which it not reliable as we have seen in Section~\ref{sec:distancebasedprioritizer}.
In contrast, performance-based methods reject or accept received models based on their performance on a test dataset.
Examples include RONI (Reject On Negative Influence) \cite{RN74} which simply discards all models with a negative impact on the model, and Zeno \cite{RN133} / Zeno++ \cite{RN102} which use a central oracle to estimate the true gradient and only keep the gradients most similar to this estimation.
Bristle uses distances only to prioritize the received models and uses per-class performance measurements to integrate models.

\textbf{Communication efficiency. }
Because of the bandwidth limitations of cellular networks, numerous methods were proposed to improve the communication efficiency of FL.
A popular method is to quantize the gradients to low-precision values \cite{RN332,RN331,RN351} or only to transmit the most important parameters (sparse matrix methods) \cite{RN334,RN347}. Bristle only updates the output layer, which works well together with existing techniques to reduce the communication overhead.

\textbf{Non-i.i.d. learning. }
Several methods enable learning on non-i.i.d. classes, such as by sharing a small i.i.d. training dataset across all peers \cite{RN254} or by reusing non-federated continual learning techniques \cite{RN256, RN301, RN303, RN304}.
Bristle also lends several concepts from a non-federated continual learning technique, namely CWR* \cite{RN323} (see Section~\ref{sec:performancebasedintegrator}).

\textbf{FL systems. }
Several FL systems try to combine Byzantine-resilience with learning on non-i.i.d. data.
RSA \cite{RN190} uses a regularization-based strategy and although it performs relatively well when the data is non-i.i.d., it fares poorly against Byzantine attackers.
FoolsGold \cite{RN40} detects and rejects attacks executed by multiple sybils working together and works well with non-i.i.d. data.
However, Zhao et al. \cite{RN138} show that the Byzantine-resilience of FoolsGold is quite limited.
FLeet \cite{damaskinos2020fleet} uses past observed staleness values and similarities with past learning tasks to achieve learning on non-i.i.d. data for a specific type of "soft" Byzantine attacks (namely the presence of stale models) but is unsuitable for other types.

\section{Concluding Remarks}\label{sec:conclusion}
We have presented Bristle, middleware for decentralized, federated learning that tolerates Byzantine attacks, even when the classes of the training data are non-i.i.d.
By leveraging deep transfer learning, Bristle achieves a high convergence rate despite having a low communication overhead.
Through the combination of a fast distance-based prioritizer and a per-class performance integrator, Bristle is able to withstand attacks targeted at subverting the model accuracy.
Our experimental evaluation using the popular MNIST dataset has demonstrated these desirable properties and shows that Bristle exhibits superior performance compared to related solutions.
In this evaluation, we did not focus on privacy, communication compression, or other aspects that are supplementary to Bristle and addressed in the existing literature.
We have implemented and open-sourced Bristle on GitHub.

Although we believe that Bristle is a significant step forward for DFL, we identify two directions for further work.
First, communication costs could be further reduced by selectively sending parameters to peers with high class overlap. This can be estimated using Private Set Intersection Cardinality (PSI-CA).
One can prevent peers from learning too much about the class distribution of others, e.g., by adding noise to the cardinality and by rate-limiting PSI-CA requests.
Second, the accuracy could be increased even more when the non-output layers can also be fine-tuned rather than being fixed.
Several popular continual learning algorithms are suitable for this purpose, such as LWF \cite{RN292}, AR1 \cite{RN323}, or EWC \cite{kirkpatrick2017overcoming}.
However, tuning also the non-output layers requires the transmission of information about these non-output layers, significantly increasing the communication costs.
Additionally, it is non-trivial to maintain the Byzantine-resilience property because the parameters in the hidden layers do not correspond directly with the classes to be predicted.

\bibliographystyle{ACM-Reference-Format}
\bibliography{paper}

\end{document}